\documentclass[aip,graphicx]{revtex4}

\usepackage{graphicx}
\usepackage{amssymb}
\usepackage{amsmath}
\usepackage{amsfonts}

\begin{document}
\title{Polymer translocation under time-dependent driving forces: resonant activation induced by attractive polymer-pore interactions}

\author{Timo Ikonen}
\affiliation{Department of Applied Physics, Aalto University School of Science,
P.O. Box 11000,
FI-00076 Aalto, Espoo, Finland}

\author{Jaeoh Shin}
\affiliation{Department of Physics, Pohang University of Science and Technology, Pohang 790-784, South Korea}

\author{Tapio Ala-Nissila}
\affiliation{Department of Applied Physics, Aalto University School of Science,
P.O. Box 11000,
FI-00076 Aalto, Espoo, Finland}
\affiliation{Department of Physics,
Box 1843, Brown University, Providence, Rhode Island 02912-1843}

\author{Wokyung Sung}
\affiliation{Department of Physics, Pohang University of Science and Technology, Pohang 790-784, South Korea}

\date{April 13, 2012}
\begin{abstract}

We study the driven translocation of polymers under time-dependent driving forces using $N$-particle Langevin dynamics simulations. We consider the force to be either sinusoidally oscillating in time or dichotomic noise with exponential correlation time, to mimic both plausible experimental setups and naturally occurring biological conditions. In addition, we consider both the case of purely repulsive polymer-pore interactions and the case with additional attractive polymer-pore interactions, typically occurring inside biological pores. We find that the nature of the interaction fundamentally affects the translocation dynamics. For the non-attractive pore, the translocation time crosses over to a fast translocation regime as the frequency of the driving force decreases. In the attractive pore case, because of a free energy well induced inside the pore, the translocation time can be a minimum at the optimal frequency of the force, the so-called resonant activation. In the latter case, we examine the effect of various physical parameters on the resonant activation, and explain our observations using simple theoretical arguments. 
\end{abstract}

\maketitle

\section{Introduction}
\label{sec:intro}

Translocation of polymers across a nanopore is a ubiquitous process in biology, with examples such as DNA and RNA transport through nuclear pore complex, protein transport through membrane channels, and virus injection into cells~\cite{albert}. Kasianowicz \textit{et al}.~\cite{Kasianowicz} demonstrated \textit{in vitro} that an electric field can transport single-stranded (ss) nucleotides through an $\alpha$-hemolysin membrane channel and it is possible to characterize individual molecules by measuring the ionic current blockade when the chain moves through the pore. Later, Li \textit{et al}. showed~\cite{li2001} that also solid-state nanopores can be used for similar experiments with a tunable size of the pore. To further the understanding of numerous biological processes and examine the perspective of technological applications such as sequencing and gene therapy, there have been extensive experimental~\cite{akeson1999, meller2000, meller2001, meller2002, meller2003, storm2005, branton2008, zwolak2008} and theoretical studies~\cite{sung1996, muthukumar1999, chuang2001, muthukumar2003, kantor2004, matysiak2006, luo2006a, luo2006b, huopaniemi2006, huopaniemi2007, luo2007a, luo2007b, luo2008a, luo2008b, luo2008c, luo2008d,dubbeldam2007a, dubbeldam2007b, panja2007, panja2008, vocks2008,milchev2011,sakaue2007,sakaue2010,saito2011,ikonen2011}. 

One of the most important quantities of the process is the translocation time and its dependence on the various system parameters such as chain length, type of driving force, pore width, etc. Even with the same chain lengths, recent experiments~\cite{akeson1999, meller2000, meller2001, meller2002} have shown that different nucleotides exhibit unique patterns in, e.g., the translocation time distribution. In particular, Meller \textit{et al}.~\cite{meller2000,meller2002} have shown that in the translocation can discriminate between polydeoxyadenylic acid (poly(dA)) and polydeoxycytidylic acid (poly(dC)) with the same chain length. The translocation time of poly(dA) is found to be longer with an exponential distribution while that of poly(dC) is shorter with a narrow distribution. The origin of the different behavior for each nucleotide was attributed to different interaction between the polymer and the pore. Recent simulation studies of Luo {\it et al.}~\cite{luo2007a, luo2008b} quantitatively support this idea.
    
Until now, most of  the \textit{in vitro} experimental as well as theoretical studies of polymer translocation have used static driving forces. However, it could be important to consider time-dependent forces to understand the process \textit{in vivo}. In a cellular environment the driving forces can be time-dependent due to the nonequilibrium fluctuations in the membrane potential, fluid density, and ionic strength, etc. In the case of translocation driven by a molecular motor~\cite{smith2001}, depending on the ATP concentration the driving force can also fluctuate. Motivated by these facts, Park and Sung~\cite{park1998} studied the translocation of a rigid rod in the presence of a dichotomically fluctuating force. They found that the system exhibits resonant activation~\cite{doering1992}, where the translocation time attains a minimum at an optimum flipping rate of the dichotomic force that is comparable to the translocation rate in the absence of the force. Although the study gives valuable insight on the effects of fluctuating forces in polymer translocation, the study is somewhat limited, however, as the flexibility of the chain is not considered and, the reflecting boundary condition which forbids the chain escape to the $cis$ side is in many cases artificial. Recent molecular dynamics simulations~\cite{sigalov2008} have shown that an alternating electric field in a nanopore exhibits a unique hysteresis in the nucleotide's dipole moment and in the chains back-and-forth motion arising from the reorientation of the DNA bases in the nanopore constriction. The authors suggest detection of DNA sequences by measuring the potential or the change of the DNA mobility in the pore. This study indicates that a time-dependent driving force may be useful for technological application as well. In addition, recent Langevin dynamics study shows that the translocation time can be significantly shortened by oscillations of the pore width~\cite{cohen2011}, which accentuates the importance of polymer-pore interactions in the problem. 

In addition to polymer translocation, there are a few simulation studies of different types of polymer transport in the presence of time-dependent driving forces. Tessier and Slater~\cite{tessier2001} considered polymer transport through a microchannel in the presence of a periodic driving force, where they found that the mobility can have a maximum at an optimal frequency. More recently, Pizzolato \textit{et al}.~\cite{pizzolato2010} have studied the effects of sinusoidal driving force on the polymer barrier crossing over a metastable potential, which is also subject to the reflecting boundary condition. They found a similar resonant behavior of the barrier crossing time. In addition, Fiasconaro \textit{et al.} have studied the one-dimensional (1D) polymer chain in the presence of sinusoidal~\cite{fiasconaro2010} and dichotomically fluctuating~\cite{fiasconaro2011} driving forces. They found that the sinusoidal driving force may induce an oscillating behavior of the translocation time~\cite{fiasconaro2010}, whereas the dichotomic force does not~\cite{fiasconaro2011}.
 
Despite the related work found in the literature, the effect of the polymer-pore interactions on the translocation of biopolymers under time-dependent driving forces and in a realistic geometry needs to be studied. In this work, the effects of time-dependent driving forces on the translocation dynamics are investigated as a first step towards understanding translocation both \textit{in vivo} and in practical applications. We consider both dichotomically fluctuating forces as an example of \textit{in vivo} nonequilibrium noise~\cite{park1998, shin2012} and sinusoidal driving forces, which might be easier to implement experimentally. We find that the polymer-pore attraction fundamentally changes the behavior of the translocation time with respect to the flipping rate of the dichotomic force or the angular frequency of the sinusoidal force. For the non-attractive pore, the translocation time has a cross-over to a fast translocation regime at low flipping rates (frequencies), but does not have a resonant minimum. For the attractive pore, we show that the system exhibits resonant activation within a broad range of physical parameters. We examine the effect of parameters such as chain length, driving force and polymer-pore interaction strength on the resonance. The results suggest that \textit{in vitro} experiments with time-dependent driving force might be useful to DNA sequencing.

\section{Model and Method}
\label{sec:model}

\begin{figure}
\includegraphics[bb=0.0cm 0.0cm 8.5cm 8.0cm]{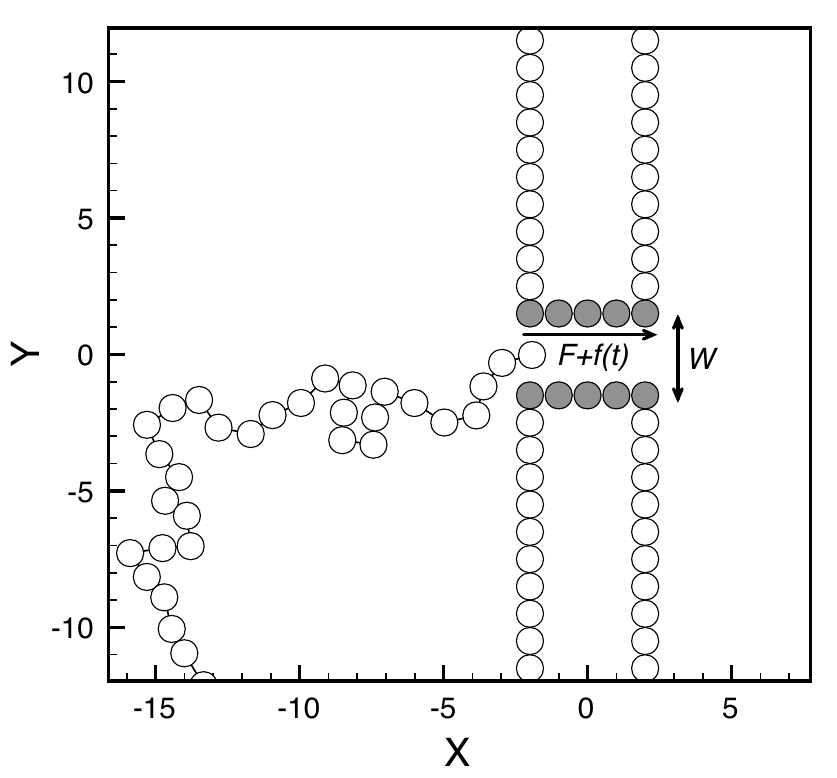}
\caption{A schematic representation of the system. The polymer, placed initially on the {\it cis} side, is driven through the pore of length $L=5$ and width $W=3$ by the time-dependent external force $F+f(t)$.\label{fig:geometry}}
\end{figure}

We consider the translocation of a self-avoiding chain in two dimensions (2D). The polymer chain is modeled by Lennard-Jones particles interconnected by finitely extensible nonlinear elastic (FENE) springs. Excluded volume interaction between monomers is given by the short-range repulsive Lennard-Jones potential: $U_\mathrm{LJ}(r)=4\epsilon \left[ \left(\frac{\sigma}{r}\right)^{12} - \left(\frac{\sigma}{r}\right)^{6}\right] +\epsilon$  for $r\leq 2^{1/6}\sigma$ and $0$ for $r>2^{1/6}\sigma$. Here, $r$ is the distance between monomers, $\sigma$ is the diameter of the monomer and $\epsilon$ is the depth of the potential well. Neighboring monomers are also connected by FENE springs with $U_\mathrm{FENE}(r)=-\frac{1}{2}kR_0^2\ln (1-r^2/R_0^2)$, where $k$ is the FENE spring constant and $R_0$ is the maximum allowed separation between consecutive monomers. The geometry of the system is shown in Fig.~\ref{fig:geometry}. The wall is constructed of immobile Lennard-Jones beads of size $\sigma$. All monomer-wall particle pairs have the same short-range repulsive LJ interaction as described above. To investigate the effect of polymer-pore interactions, we consider two main types of interactions between the monomers and the pore particles: attractive and non-attractive. In the case of non-attractive interactions, the pore particles are considered to be identical with the wall particles, having a purely repulsive interaction with the monomers. In the case of the attractive polymer-pore interactions, the cut-off distance of the LJ potential between monomer-pore particles is increased to $2.5\sigma$ (with $U_\mathrm{LJ}$ constant for $r>2.5\sigma$), and the interaction strength is characterized by $\epsilon_\mathrm{pm}$.  The interaction can be either attractive or repulsive, depending on the distance of the monomer from the pore particles.

In our simulations, the dynamics of each monomer is described by the Langevin equation
\begin{equation}
m \mathbf{\ddot{r}}_i=-\nabla (U_\mathrm{LJ}+U_\mathrm{FENE})+\mathbf{F}_\mathrm{ext}-\xi \mathbf{v}_i +\mathbf{F}^R_i\label{eq:eqmotion},
\end{equation}
 where $m$ is the monomer mass, $\xi$ is the friction coefficient, $\mathbf{v}_i$ is the monomer velocity and $\mathbf{F}^R_i$ is the random force with correlations $\langle  \mathbf{F}^R_i(t) \cdot \mathbf{F}^R_j(t') \rangle = 4\xi k_BT \delta_{i,j}\delta(t-t')$, where $k_B$ is the Boltzmann constant and $T$ is the temperature. In the pore, the monomers experience an external driving force $\mathbf{F}_\mathrm{ext}=[F+f(t)]\hat{x}$, where $F$ is static (time-independent) force, $f(t)$ is the time-dependent force and $\hat{x}$ is the unit vector along the direction of the pore axis. In this work, we consider two types of time-dependent forces $f(t)$. The first is the dichotomic noise, for which $f(t)$ is either $+A_{d}$ or $-A_{d}$, and changes from one value to the other with flipping rate $\omega$. The dichotomic $f(t)$ has zero mean and is exponentially correlated: $\langle f(t)\rangle=0$ and $\langle f(t)f(0)\rangle=A_{d}^2\exp(-2\omega t)$. As a second example, we consider the sinusoidal force given by $f(t)=A\sin(\omega t + \phi)$, where $A$ is the amplitude, $\omega$ the angular frequency and $\phi$ is a constant phase.

We use the LJ parameters $\epsilon$, $\sigma$ and $m$ to fix the scales for energy, length and mass, respectively. The time scale is then given by $t_\mathrm{LJ}=(m\sigma^2/\epsilon)^{1/2}$. The dimensionless parameters in our simulations are $R_0=2$, $k=7$, $\xi=0.7$ and $k_BT=1.2$. In our model, the bead size corresponds to the Kuhn length of a single-strand DNA, giving approximately $\sigma\approx 1.5$~nm. The bead mass is approximately 936 amu, and the interaction strength $\epsilon$ corresponds to $3.39\cdot 10^{-21}$~J at room temperature (295 K). The Lennard-Jones time scale is then $32.1$~ps. Where appropriate, we will express our results also in terms of the mean translocation time in the absence of $f(t)$, $\tau_0$, which is the physically relevant time scale in the system and also easy to measure experimentally. The pore dimensions we set as $L=5$ and $W=3$, as shown in Fig.~\ref{fig:geometry}. With the force scale of 2.3 pN, a static driving force of $F_\mathrm{ext}=1$ then corresponds to a voltage of 375 mV across the pore (assuming three unit charges per bead and the effective charge $0.094e$ for a unit charge~\cite{sauer2003}). The equations of motion are integrated with the Ermak algorithm \cite{ermak, allen} with time-step typically $\Delta t=0.01$, and shorter when necessary.

Initially, the first monomer of the chain is held fixed at the pore (see Fig.~\ref{fig:geometry}) while the remaining monomers are allowed to fluctuate until an equilibrium configuration is reached. Then at time $t=0$ the first monomer is released and the external force is applied.  For the dichotomic force, the initial force is randomly selected from $+A_d$ and $-A_d$ with equal probability. Correspondingly, for the sinusoidal force, the phase $\phi$ is randomly selected from a uniform distribution between $[0,2\pi]$. For small $\mathbf{F}_\mathrm{ext}$ and weak polymer-pore attraction, the chain may slip out of the pore back to the {\it cis} side instead of translocating to the {\it trans} side. In that case, the equilibration process is repeated and the simulation is begun anew.  The process is repeated until at least 2000 successful translocation events are recorded. In addition to this standard procedure, it is possible to impose a reflecting boundary condition that prevents the first bead from slipping back to the {\it cis} side. In this case, the simulation is run simply until a successful translocation occurs. It turns out that this boundary condition, although widely used in translocation study, fundamentally changes the translocation dynamics, as will be discussed in Section~\ref{sec:results}. That is why, unless otherwise indicated, all the results presented in this work have been computed {\it without} the reflecting boundary condition.

\section{Results and discussion}
\label{sec:results}

\subsection{Non-attractive pore, dichotomic driving force}
\label{sec:nonattr_dichotomic}

We begin by considering the purely repulsive polymer-pore interactions, which is the most common case studied in the literature. The strength of the Lennard-Jones interaction is $\epsilon_\mathrm{pm}=1$ with a cut-off distance of $2^{1/6}\sigma$. First, we consider the dichotomic driving force, with the results for the sinusoidal force presented later in Sec.~\ref{sec:nonattr_periodic}. We have chosen the numerical values $F=0.3$ and $A_d=0.2$ for the dichotomic force and $N=64$ for the chain length, which are within the experimental regime. We have checked that within the experimentally relevant force regime and at least for $N \leq 128$ the qualitative behavior remains the same. 

\begin{figure}
\includegraphics[bb=0.0cm 0.0cm 8.5cm 5.6cm]{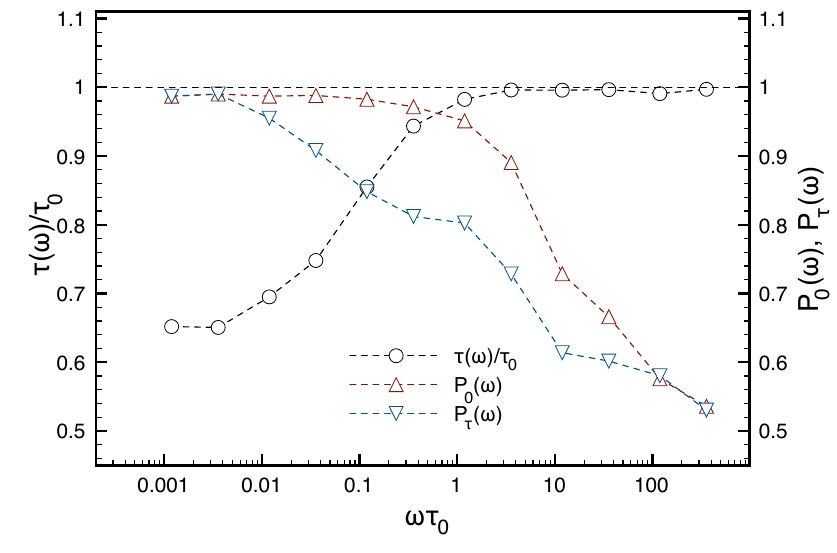}
\caption{The mean translocation time $\tau$ and the probabilities $P_0$ and $P_\tau$ (see text) as a function of the flipping rate $\omega$ of the dichotomic force for the repulsive pore. $N=64$, $F=0.3$, $A_d=0.2$ and $\tau_0\approx 750\pm4$. The statistical error is smaller than the symbol size. \label{fig:nonattr_dichotomic_tau_p0_ptau}}
\end{figure}

The main results for the dichotomic force as a function of the flipping rate $\omega$ are gathered in Fig.~\ref{fig:nonattr_dichotomic_tau_p0_ptau}. As a function of the flipping rate $\omega$, we observe two distinct regions. In the fast flipping regime, $\omega\gg 1/\tau_{0}$, the average translocation time is $\tau(\omega)\approx\tau_0$. Here, $\tau_0$ is the translocation time in the absence of dichotomic forces, i.e., $A_d=0$. In this limit,  due to the high flipping rate, $f(t)$ changes its sign many times during the course of the translocation and is averaged out to zero over the whole process. Therefore, we have $\tau(\omega)\approx\tau_0$ for $\omega\rightarrow\infty$. In principle, in this limit, the time-dependent force becomes a rapidly fluctuating $\delta$-correlated noise similar to the thermal random force $\mathbf{F}_i^\mathrm{R}$. For the monomers inside the pore, the modified correlation of the random force is given as $\langle [ f(t) \hat{x}+\mathbf{F}^\mathrm{R}(t)] \cdot [ f(0) \hat{x}+\mathbf{F}^\mathrm{R}(0)] \rangle = (4\xi k_BT + A_d^2/\omega)\delta(t)= 4\xi k_BT \delta(t)$, where last expression is obtained in the limit $\omega\rightarrow\infty$. Therefore, the effect of the dichotomic force in this limit is vanishing, and we recover $\tau(\omega)\approx\tau_0$, as shown in Fig.~\ref{fig:nonattr_dichotomic_tau_p0_ptau}. This result is also in agreement with Refs.~\cite{pizzolato2010,fiasconaro2010,fiasconaro2011}. As the flipping becomes slower, $\omega < 1/\tau_0$, we observe a cross-over to a faster translocation regime, with $\tau(\omega)<\tau_0$. This result is in sharp contrast with Refs.~\cite{pizzolato2010,fiasconaro2011,fiasconaro2010}, where it was found that $\tau(\omega)>\tau_0$. In addition, we do not find a global minimum of $\tau(\omega)$ at any finite $\omega$, unlike Refs.~\cite{pizzolato2010,fiasconaro2010}.

\begin{figure}
\includegraphics[bb=0.0cm 0.0cm 8.5cm 6.0cm]{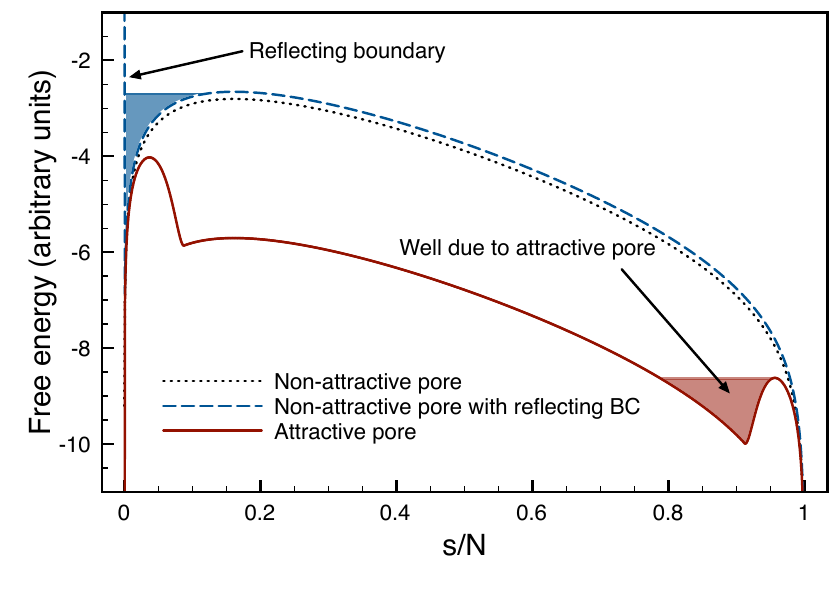}
\caption{Illustration of the free energy of the polymer chain as a function of the number of translocated monomers $s$. The dotted line indicates the free energy for the non-attractive pore, which has no well structure. A reflecting boundary condition at $s=0$ forms a free-energy well (blue shaded area). Attractive polymer-pore interactions can also create a free-energy well (schematically shaded red).  \label{fig:free_energy_schema}}
\end{figure}

To understand the behavior of $\tau(\omega)$ at small $\omega$, we need to look at the probability of achieving a successful translocation. Due to confinement within the pore, the chain experiences an entropic free energy barrier~\cite{sung1996, muthukumar1999,muthukumar2003}, as illustrated in Fig.~\ref{fig:free_energy_schema}. Because of fluctuations, there is a finite probability that the chain slips back to the {\it cis} side instead of translocating to the {\it trans} side. Therefore, the probability of translocation is less than one and increases with increasing driving force (for details of the translocation probability as a function of various system parameters, see Ref.~\cite{luo2007a}). Thus, within the set of successful translocations, we expect to find a larger number events that have positive $f(t)$, as compared to those with negative $f(t)$. We characterize this dependence of the translocation probability on the driving force by looking at the set of successful translocations, from which we calculate the distribution of $f(t)$ at the beginning of translocation ($t=0$) and at the final moment of translocation ($t=\tau$). The probabilities $P_0 \equiv P[f(0)>0]$ and $P_\tau \equiv P[f(\tau)>0]$ that the force $f(t)$ is positive for $t=0$ and $t=\tau$, respectively, are shown in  Fig.~\ref{fig:nonattr_dichotomic_tau_p0_ptau}. In the high-rate regime, the flipping rate is too high for $f(0)$ or $f(\tau)$ to be correlated with the chain dynamics, and therefore $P_0$ and $P_\tau$ approach 0.5. On the other hand, in the low-rate regime, the positive direction of $f(0)$ is strongly favored ($P_0\approx 0.98$). In addition, since the correlation time of the driving force is much longer than $\tau$, the driving force remains constant during the whole translocation process with high probability, being either $F_\mathrm{ext}=F+A_d$ or $F_\mathrm{ext}=F-A_d$. Therefore, in this limit, the average translocation time is given by the weighted average
\begin{equation}
\tau= P_0\tau_+ + (1-P_0)\tau_-.\label{eq:tau_ave_dichotomic}
\end{equation}
Here, $\tau_{+}$ and $\tau_{-}$ are the translocation times with the total force $F+A_{d}$ and $F-A_{d}$, respectively. Assuming that the translocation time is inversely proportional to the driving force, $\tau(f)\sim f^{-1}$, Eq.~(\ref{eq:tau_ave_dichotomic}) gives $\tau(\omega)\approx 0.65\tau_0$ in the low-$\omega$ limit. This agrees well with the results in Fig.~\ref{fig:nonattr_dichotomic_tau_p0_ptau}. Therefore, the cross-over to the fast translocation regime ($\tau < \tau_0$) as the flipping becomes slower is simply explained by the fact that for low flipping rate the chain is most likely to translocate when $f(0)>0$. This strong bias for selecting the initial value $f(0)$ induced by the entropic barrier is the crucial difference between this work and Refs.~\cite{pizzolato2010,fiasconaro2010,fiasconaro2011} In Refs.~\cite{pizzolato2010,fiasconaro2010,fiasconaro2011} this kind of selection does not occur because the translocation probability is one, independent of the time-dependent driving force. A similar effect can be obtained in our model by imposing a reflecting boundary condition that prevents the first monomer from exiting the pore to the {\it cis} side. However, we stress that this kind of boundary condition may not be realistic for, e.g., the translocation of a ss-DNA molecule through a pore, although it has been used in many studies.

\begin{figure}
\includegraphics[bb=0.0cm 0.0cm 8.5cm 4.5cm]{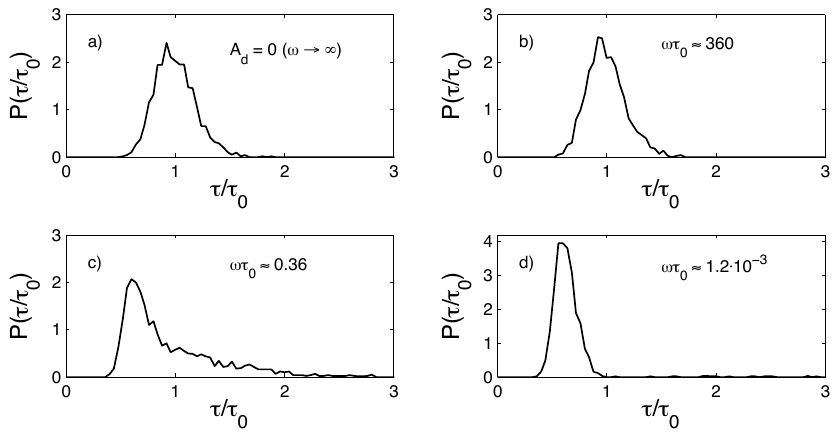}
\caption{The distribution of translocation times for chain length $N=64$ and $F=0.3$ under dichotomic driving force in the non-attractive pore. Panel a) shows the distribution for $A_d=0$, while panels b) -- d) show the distribution for $A_d=0.2$. \label{fig:nonattr_dichotomic_histograms}}
\end{figure}

Finally, we look at the distribution of translocation times. In the high flipping rate limit, the distribution is very similar to the zero amplitude case (see. Fig.~\ref{fig:nonattr_dichotomic_histograms}). In general, the distribution at this limit is either Gaussian (for large enough $F$) or has an exponentially decaying tail (for small $F$). In the present case, the distribution is almost Gaussian with a slightly elongated tail, which differs greatly from the typical distributions of thermally activated processes. Furthermore, as shown in Fig.~\ref{fig:free_energy_schema}, there is no metastable well (pretransition state) within which the chain attempting to escape would oscillate. Thus, in this case, the resonant minimum of $\tau(\omega)$ does not exist, in contrast to Ref.~\cite{pizzolato2010}, where the adopted external potential has a pretransitional well. At lower flipping rates, the peak of the distribution moves toward shorter translocation times, as the trajectories having $f(t)$ predominantly in the positive direction are favored (signaled by increasing $P_0$ and $P_\tau$). However, also the probability of long translocation times increases. These events correspond to the trajectories with negative $f(t)$. As the flipping rate is further decreased, most of the successful translocations occur with $f(t)>0$. In the low-rate limit, one retains two peaks, corresponding to $F_\mathrm{ext}=F+A_d$ and $F_\mathrm{ext}=F-A_d$. For $F=0.3$ and $A_d=0.2$, only the former is practically visible.

\subsection{Non-attractive pore, periodic driving force}
\label{sec:nonattr_periodic}

\begin{figure}
\includegraphics[bb=0.0cm 0.0cm 8.5cm 5.6cm]{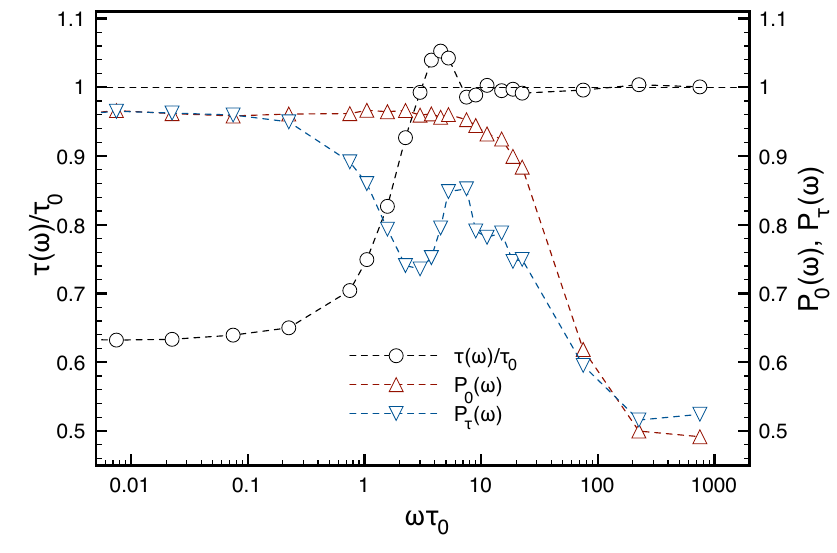}
\caption{The mean translocation time $\tau$ and the probabilities $P_0$ and $P_\tau$ as a function of the angular frequency $\omega$ for the periodic force and repulsive pore. $N=64$, $F=A=0.3$ and $\tau_0\approx 750\pm4$. The statistical error is smaller than the symbol size.  \label{fig:nonattr_tau_p0_ptau}}
\end{figure}

As a second case, we study the translocation through a non-attractive pore under sinusoidally time-dependent driving force $f(t)=A\sin(\omega t + \phi)$. The average translocation time $\tau(\omega)$ and the probabilities $P_0$ and $P_\tau$ are shown in Fig.~\ref{fig:nonattr_tau_p0_ptau} as a function of the angular frequency $\omega$. For comparison with the dichotomic case, we use the parameter values $N=64$, $F=0.3$ and $A=0.3$. The time-averaged amplitude of the time-dependent force is then $\langle | A\sin(\omega t) | \rangle_t=2A/\pi\approx 0.2$, which corresponds to the value of $A_d$ used in the previous Section. In the low-frequency ($\omega \ll 1/\tau_0$) and high-frequency ($\omega \gg 1/\tau_0$) limits we obtain results similar to the dichotomic force explained above: in the high-$\omega$ limit, $\tau(\omega)\approx \tau_0$ and, in the opposite limit of small $\omega$, $\tau(\omega)<\tau_0$. The average translocation time is given by a relation analogous to Eq.~(\ref{eq:tau_ave_dichotomic}):
\begin{equation}
\tau=\frac{1}{2\pi}\int_0^{2\pi}p(\phi)\tau(\phi)d\phi.\label{eq:tau_ave}
\end{equation}
Here $p(\phi)$ is the probability density of the initial phase $\phi$ within the set of successful translocations and $\tau(\phi)$ is the average translocation time corresponding to the driving force $F+A\sin(\phi)$. Similarly to the dichotomic case, the distribution $p(\phi)$ is uniform only in the high frequency limit, while in the low frequency limit, values of $\phi$ giving $f(0)>0$ are strongly favored (cf. Fig.~\ref{fig:nonattr_tau_p0_ptau}), as we shall see. This leads to larger average driving forces and consequently faster translocation.

In the intermediate regime ($\omega\approx 1/\tau_0$), the periodic time-dependence of the driving force fundamentally affects the translocation dynamics. Instead of a simple cross-over in $\tau(\omega)$, one gets a series of local minima and maxima. In addition, the probability $P_\tau$ has a local maximum in the vicinity of a local minimum of $\tau$. In many cases, these could be argued to indicate resonant activation~\cite{doering1992}. However, in this case they have a deterministic origin weighted with the distribution $p(\phi)$. To show this, we consider a coarse-grained model for the translocated segments already studied in Ref.~\cite{sung1996,muthukumar1999}, with the entropic contributions therein neglected to make it analytically solvable. This approximation is reasonable because apart from the short initial (and final) stage of translocation, the entropic force is small compared to the mean driving force $F$. Our model is the 1D equation of motion for $\langle s(t) \rangle$, the average number of translocated segments, under the sinusoidal driving force with a fixed value of $\phi$,
\begin{equation}
\xi_\mathrm{eff} \frac{d\langle s(t) \rangle}{dt} = F\left[ 1+\sin(\omega t +\phi) \right],\label{eq:toy_model}
\end{equation} 
where $\xi_\mathrm{eff}$ is the effective friction.  Eq.~(\ref{eq:toy_model}) can be analytically solved for $\langle s(t) \rangle$ with the initial condition $s(0)=0$. Specifically, we are interested in the time that it takes for the system to evolve from $s=0$ to $s=N$ as a function of the phase, $\tau(\phi)$. We fix $\xi_\mathrm{eff}$ by setting the time-scale of the model so that $F/\xi_\mathrm{eff}=N/\tau_0$, giving $\xi_\mathrm{eff}\approx 17.6$ for $N=64$ and $F=A=0.3$. The integration of Eq.~(\ref{eq:toy_model}) yields
\begin{equation}
\tau(\phi)=\tau_0 + \frac{1}{\omega}\left[ \cos(\omega \tau + \phi) - \cos(\phi) \right].\label{eq:toy_model_solution}
\end{equation}
This describes the approach of the translocation time $\tau$ to $\tau_0$ in the $\omega \rightarrow \infty$ limit, as well as the local oscillation in the intermediate $\omega$ regimes. Once $\tau(\phi)$ is obtained as a function of $\phi$ as well as $\omega$ from Eq.~(\ref{eq:toy_model_solution}), the translocation time averaged over $\phi$ is found from Eq.~(\ref{eq:tau_ave})

\begin{figure}
\includegraphics[bb=0.0cm 0.0cm 8.5cm 6.0cm]{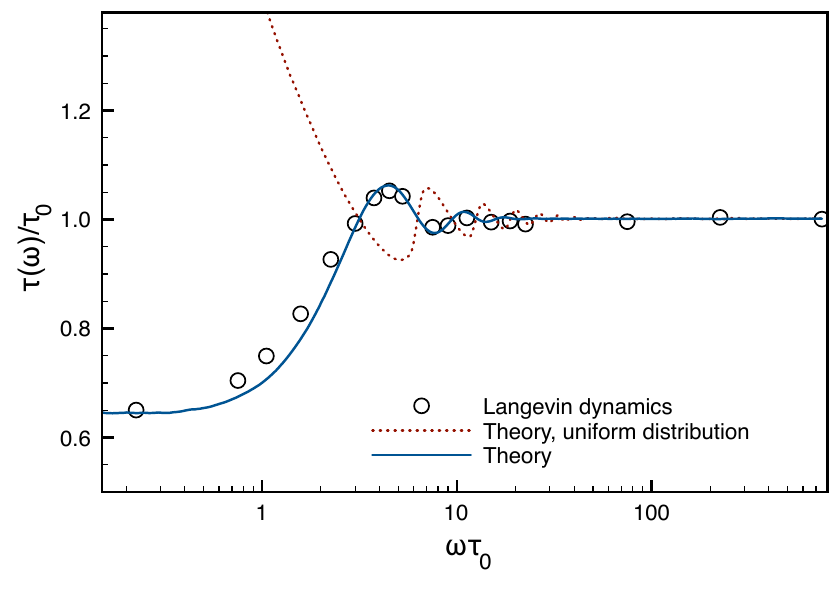}
\caption{ Comparison between LD simulations  ($N=64$, $F=A=0.3$) and the theoretical toy model. Dotted line:  toy model with uniformly distributed $\phi$, solid line: toy model with Boltzmann distributed $\phi$ (see text). The latter shows good agreement with the LD results (circles). \label{fig:nonattr_tau_different_models}}
\end{figure}

In Fig.~\ref{fig:nonattr_tau_different_models}, we compare our model with the $N$-particle Langevin dynamics simulations. First, the dotted line shows the results for a uniformly distributed $\phi$. In contrast to the LD simulations, this curve shows a global minimum of translocation time, and also a strong oscillating behavior as a function of $\omega$. The behavior is very similar to the simple 1D chain model driven by sinusoidal force studied in Ref.~\cite{fiasconaro2010}.

However, $\phi$ should not be chosen uniformly. In the properly formulated translocation problem, the chain has to overcome the initial free energy barrier, which leads to nonuniform distribution of $\phi$. In the zero-frequency limit, the translocation probability follows the Boltzmann distribution, which depends exponentially on the height of the initial free energy barrier (cf. Fig.~\ref{fig:free_energy_schema}). Hence, we put the distribution in the form $p(\phi)\sim\exp\left[ \alpha\sin(\phi)\right]$. Since we consider only the processes that complete the translocation, $\alpha$ is a nontrivial function of not only $k_BT$, $F$, $A$ but also $\omega$. In our procedure, the parameter $\alpha$ is obtained by fitting the integral $\int_0^\pi p(\phi) d\phi$ to the probability $P_0(\omega)$ for each $\omega$. The $\alpha$, obtained as $\alpha(\omega)=(15\omega + 1/2.6)^{-1}$, serves as an empirical interpolation between the Boltzmann distribution for $\omega \ll 1/\tau_0$ and the uniform distribution of $\phi$ for $\omega \gg 1/\tau_0$. This $\alpha$ then gives the distribution $p(\phi)$, over which the average of $\tau(\phi)$ is taken by Monte Carlo integration to eventually find the average translocation time $\tau$. The results of our model with this distribution of $\phi$ are shown in Fig.~\ref{fig:nonattr_tau_different_models} as a solid curve. The model reproduces the essential features of the full $N$-particle Langevin dynamics simulation: the cross-over to fast translocation as $\omega$ is decreased, and the global and local maxima of $\tau$ near $\omega \approx \pi/\tau_0$. This exercise clearly shows that the local maxima and minima are a result of deterministic dynamics and the nonuniform distribution of $\phi$, and are not indications of resonant activation.

\begin{figure}
\includegraphics[bb=0.0cm 0.0cm 8.5cm 6.6cm]{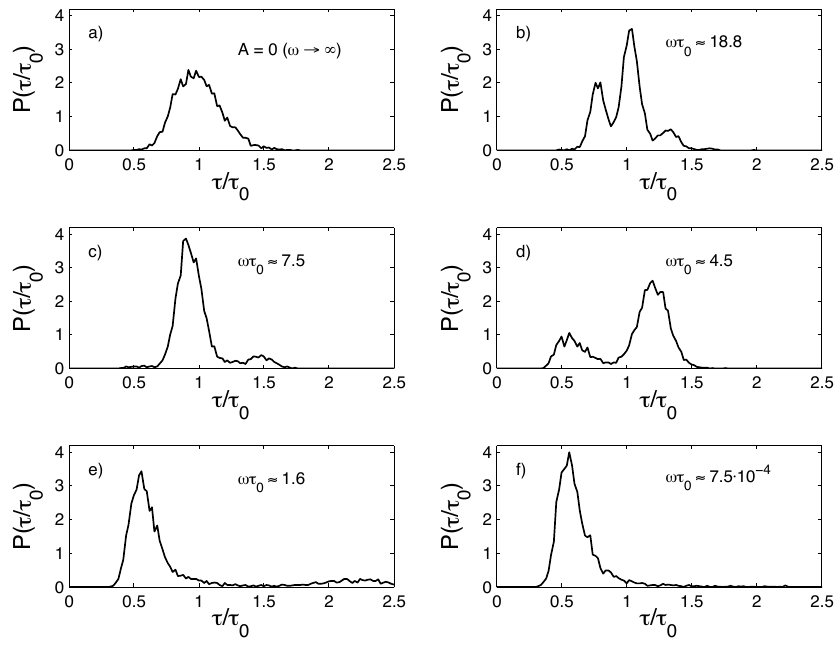}
\caption{The distribution of translocation times for chain length $N=64$ and $F=0.3$ under sinusoidal driving force in the non-attractive pore. Panel a) shows the distribution for $A=0$, while panels b) -- f) show the distribution for $A=0.3$. \label{fig:nonattr_histograms}}
\end{figure}

The difference between the sinusoidal and dichotomic driving forces can also be identified in the translocation time distributions. In the high and low frequency limits, one recovers distributions very similar to the dichotomic case. In the intermediate frequency regime, on the other hand, the sinusoidal time-dependence shows as a periodic modulation of the underlying distribution. Here, the distribution has multiple peaks, which correspond to translocations occurring when $f(t)>0$ with high probability. Each peak corresponds to one period $T_\Omega\equiv 2\pi/\omega$ of the sinusoidal force, with the distance between consecutive peaks being $\Delta\tau/\tau_0\approx 2\pi/\omega\tau_0$, as shown in Fig.~\ref{fig:nonattr_histograms} (b). Near $\omega\approx\pi/\tau_0$, where the average translocation time achieves its maximum, the distribution shows two distinct peaks, corresponding to fast ($\tau<\tau_0$) and slow ($\tau>\tau_0$) translocation (see. Fig~\ref{fig:nonattr_histograms} d). The leftmost peak corresponds to events that occur roughly between $T_\Omega/4 < \tau < T_\Omega/2$, with a phase $\phi$ between $0<\phi<\pi/2$. For these events, $f(t)$ is positive for the whole translocation process, resulting in faster than average translocation. The peak on the right, on the other hand, corresponds to the events with $3T_\Omega/4 < \tau < T_\Omega$ and $\pi/2<\phi<\pi$. Here, although $f(t)$ starts positive, it quickly crosses over to negative values. Typically, translocation occurs when $f(t)$ turns back to positive. Thus, the average $f(t)$ during one event is negative, giving longer than average translocation time. As the frequency $\omega$ is decreased, the rightmost peak becomes smaller as the phases $\phi$ corresponding to that peak become less probable. As a result, the average translocation time crosses over to the regime where $\tau<\tau_0$.

\subsection{Attractive pore, dichotomic driving force}

We have shown above that for purely repulsive pore-monomer interactions, the system does not exhibit resonant activation. This is due to the absence of a proper free-energy well, in which an attempt frequency of crossing the imminent barrier is well defined. Introducing attractive interactions between the polymer and the pore modifies the free energy in such a way that a well is formed (schematically shown in Fig.~\ref{fig:free_energy_schema}), and translocation becomes a thermally activated barrier crossing process~\cite{muthukumar2003,luo2007a, luo2008b}. Therefore, for the attractive pore, we expect to find a resonance similar to that reported for the polymer escape in Ref.~\cite{pizzolato2010}. We start with the case of dichotomic driving force, which is somewhat more pedagogical than the sinusoidal force case.

\subsubsection{Dependence on the polymer-pore interaction strength $\epsilon_{\mathrm{pm}}$}

\begin{figure}
\includegraphics[bb=0.0cm 0.0cm 8.5cm 5.5cm]{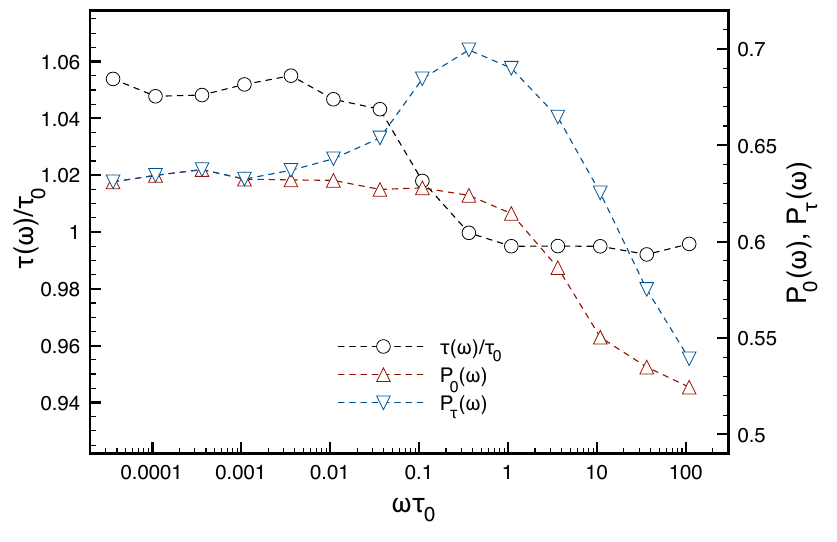}
\caption{The mean translocation time $\tau$ and the probabilities $P_0$ and $P_\tau$ for the dichotomic force and attractive pore. $N=32$, $F=0.5$, $A_{d}=0.2$, $\epsilon_\mathrm{pm}=1$, and $\tau_0\approx 226.8\pm0.6$. The statistical error is smaller than the symbol size. \label{fig:tau_probability_eps1}}
\end{figure}

\begin{figure}
\includegraphics[bb=0.0cm 0.0cm 8.5cm 5.5cm]{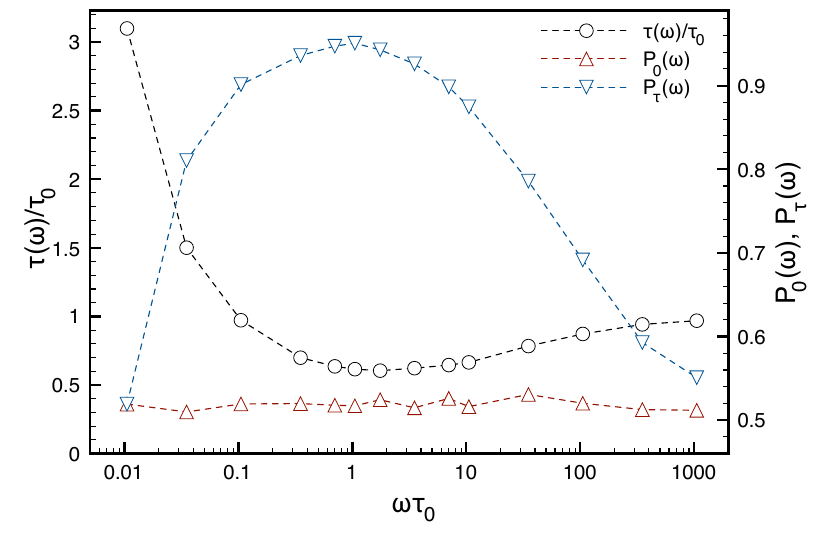}
\caption{The mean translocation time $\tau$ and the probabilities $P_0$ and $P_\tau$ for the dichotomic force and attractive pore. $N=32$, $F=0.5$, $A_{d}=0.2$, $\epsilon_\mathrm{pm}=2.5$, and $\tau_0\approx 2202\pm29$. The statistical error is smaller than the symbol size.  \label{fig:tau_probability_eps25}}
\end{figure}

First, we study the effect of the polymer-pore interaction strength $\epsilon_\mathrm{pm}$ on the average translocation time $\tau$. For the attractive pore, we use the value $2.5\sigma$ for the cut-off distance of the Lennard-Jones potential, which yields an attractive force between the pore and the monomer at distances $2^{1/6}\sigma < r < 2.5\sigma$. The chain is driven by a dichotomically fluctuating force with the flipping rate $\omega$ and correlations described in Section~\ref{sec:model}. In Figs.~\ref{fig:tau_probability_eps1} and~\ref{fig:tau_probability_eps25}, we show the average translocation times $\tau(\omega)$ for the chain length $N=32$, with the polymer-pore interaction strength $\epsilon_{\mathrm{pm}}=1$ and $\epsilon_{\mathrm{pm}}=2.5$, respectively. Here, the static force $F=0.5$ and the amplitude of dichotomic force is $A_{d}=0.2$. In the high flipping rate regime, $\omega\gg 1/\tau_{0}$, $\tau(\omega)\approx \tau_0$, as for the non-attractive pore. On the other hand, for $\omega\ll 1/\tau_{0}$, the translocation time is $\tau>\tau_0$. This behavior is completely opposite to the non-attractive pore case. Nevertheless, it can be explained by the same arguments. The average translocation time is given by Eq.~(\ref{eq:tau_ave_dichotomic}). However, for sufficiently large $\epsilon_\mathrm{pm}$ the selectivity with respect to the initial driving force $f(0)$ is fairly weak, because a strong attraction between the pore and the polymer prevents the escape to the {\it cis} side. For example, for $\epsilon_\mathrm{pm}=1.0$, $P_0\approx0.63$, as shown in Fig.~\ref{fig:tau_probability_eps1}. Assuming inverse dependence of the translocation time on the driving force, Eq.~(\ref{eq:tau_ave_dichotomic}) gives $\tau\approx 1.06\tau_0$, which is in agreement with the simulation results. 


\begin{figure}
\includegraphics[bb=0.0cm 0.0cm 8.5cm 8.1cm]{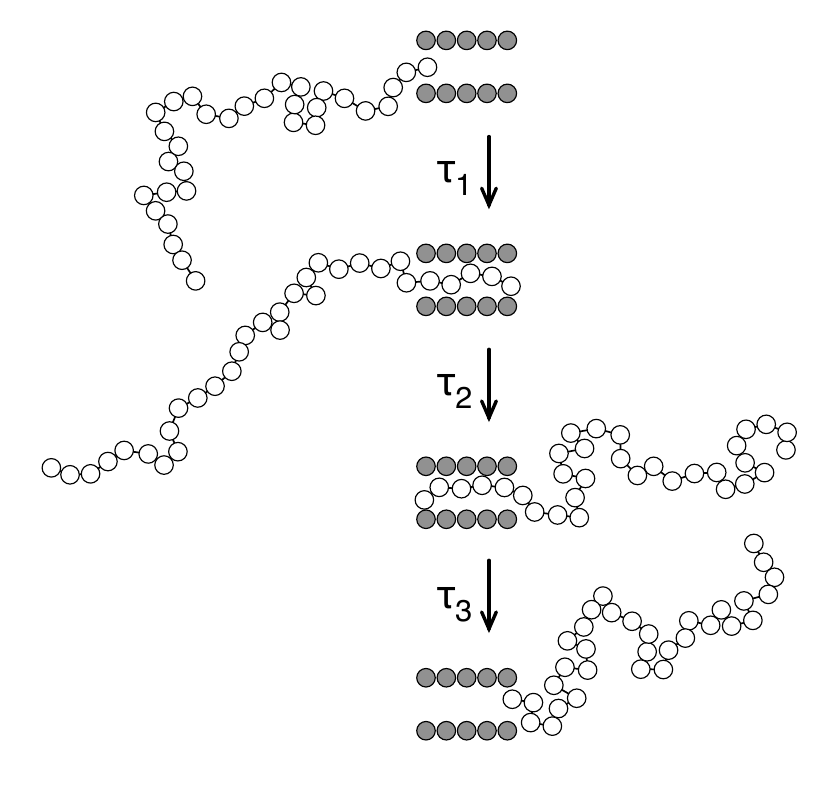}
\caption{The translocation process divided into three stages: 1) initial filling of the pore, 2) transfer of the polymer from the {\it cis} side to the {\it trans} side,  3) the final emptying of the pore. The corresponding times of the subprocesses are $\tau_1$, $\tau_2$ and $\tau_3$, with the total translocation time $\tau=\tau_1 + \tau_2 + \tau_3$.\label{fig:translocation_process}}
\end{figure}

\begin{figure}
\includegraphics[bb=0.0cm 0.0cm 8.5cm 6.0cm]{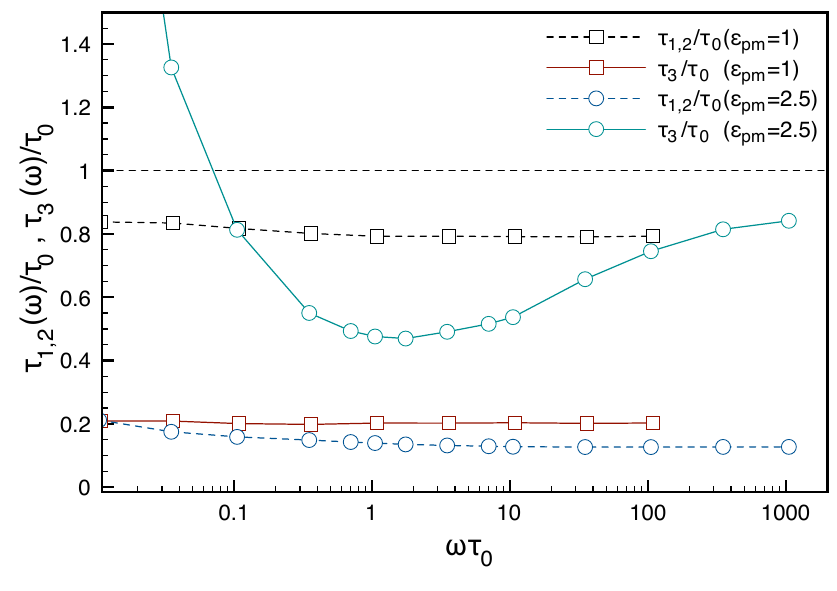}
\caption{The times $\tau_{1,2}$ (dashed lines) and $\tau_{3}$ (solid lines) for $\epsilon_\mathrm{pm}=1$ (squares) and $\epsilon_\mathrm{pm}=2.5$ (circles). Here $N=32$, $F=0.5$, $A_{d}=0.2$ for both cases. While $\tau_{1,2}$ monotonically decreases as $\omega$ increases, $\tau_{3}$ shows a resonant minimum.  The statistical error is smaller than the symbol size. \label{fig:tau_123}}
\end{figure}

In the intermediate regime ($\omega\approx 1/\tau_0$), the translocation time $\tau(\omega)$ shows different behavior depending on the value of $\epsilon_{\mathrm{pm}}$. While for $\epsilon_{\mathrm{pm}}=1$, $\tau(\omega)$ monotonically decreases as $\omega$ increases, for $\epsilon_{\mathrm{pm}}=2.5$, $\tau(\omega)$ has a minimum at an optimal flipping rate $\omega\tau_0\approx 1.8$. Related to this, we obtain the probabilities $P_{0}(\omega)$ and $P_{\tau}(\omega)$. For $\epsilon_{\mathrm{pm}}=1$, $P_{0}$ monotonically increases as $\omega$ decreases, similarly to the non-attractive case, but only by approximately $0.1$. For $\epsilon_{\mathrm{pm}}=2.5$, $P_{0}\approx0.52$, almost independent of $\omega$. On the other hand, $P_{\tau}$ shows nonmonotonic behavior, having a maximum at $\omega\tau_0\approx 0.4$  and $\omega\tau_0\approx 1.0$ for $\epsilon_{\mathrm{pm}}=1$ and 2.5, respectively. Typically, for a barrier crossing problem, such a maximum is an indication of resonant activation, and is accompanied by a minimum in the crossing time~\cite{doering1992,shin2012}. However, out of the two cases, $\epsilon_{\mathrm{pm}}=1$ and $\epsilon_{\mathrm{pm}}=2.5$, only in the latter has a minimum in $\tau(\omega)$. Furthermore, for $\epsilon_{\mathrm{pm}}=2.5$, the flipping rates $\omega$ at the maximum of $P_{\tau}(\omega)$ and at the minimum of $\tau(\omega)$ do not coincide. To understand these results, we divide the translocation process into three components~\cite{muthukumar2003,luo2007a,luo2008b}: 1) initial filling of the pore, 2) transfer of the polymer from $cis$ to $trans$ side, and 3) emptying of the pore, as shown in Fig.~\ref{fig:translocation_process}. The translocation time is then $\tau=\tau_{1}+\tau_{2}+\tau_{3}$, where $\tau_{i}$ is the time for the $i$th process. In Fig.~\ref{fig:tau_123} we show $\tau_{1,2}(\equiv\tau_{1}+\tau_{2})$ and $\tau_{3}$ for $\epsilon_\mathrm{pm}=1$ and $\epsilon_\mathrm{pm}=2.5$. For the larger $\epsilon_\mathrm{pm}$, $\tau_{3}$ dominates the translocation time. As $\omega$ increases, $\tau_{1,2}(\omega)$ decreases gradually, but $\tau_{3}(\omega)$ behaves non-monotonically. In addition, the minimum of $\tau_{3}$ coincides with the maximum of $P_{\tau}$ (for $\epsilon_{\mathrm{pm}}=1$ this is barely observable). This indicates that $P_{\tau}(\omega)$ and $\tau_{3}(\omega)$ are highly correlated. Thus, the nonmonotonic behavior of $\tau(\omega)$ occurs because the time-dependent force couples to the pore emptying process, i.e., the crossing of the final free-energy barrier (cf. Fig.~\ref{fig:free_energy_schema}). The coupling to the first two processes is very weak, and does not significantly contribute to the resonant activation. However, since $\tau_{1,2}$ slightly decreases as $\omega$ increases, the optimal flipping rate that yields the minimum of translocation time $\tau$ is somewhat larger than the rate at the minimum of $\tau_{3}$. 

\begin{figure}
\includegraphics[bb=0.0cm 0.0cm 8.5cm 7.1cm]{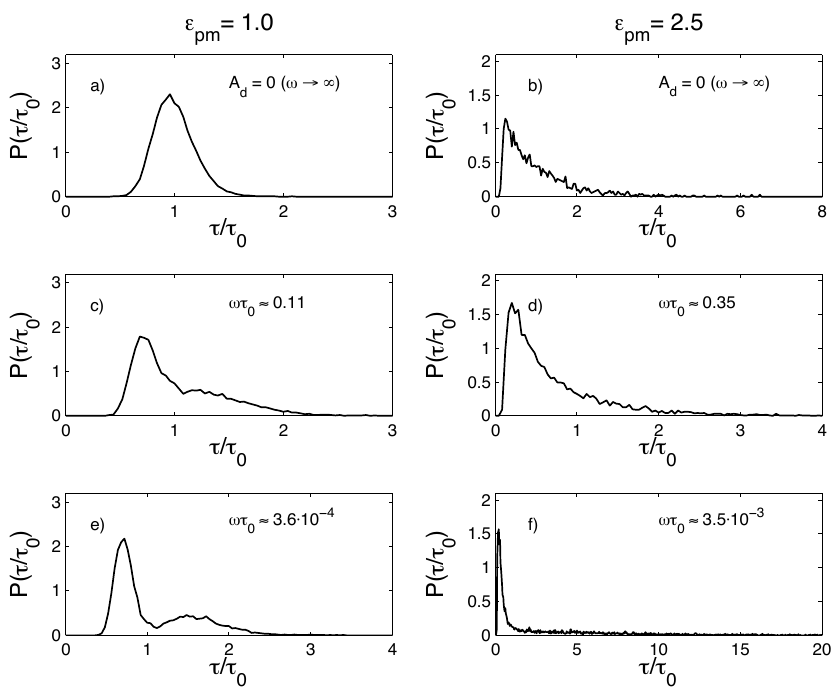}
\caption{The distribution of translocation times for the dichotomic force and attractive pore. $N=32$, $F=0.5$, and $A_d=0.2$. The left column shows the distributions for $\epsilon_\mathrm{pm}=1$ and the right column for $\epsilon_\mathrm{pm}=2.5$.
\label{fig:attraction_histogram}}
\end{figure}

The translocation time distribution $P(\tau)$ also profoundly depends on the magnitude of $\epsilon_{\mathrm{pm}}$, as shown in Fig.~\ref{fig:attraction_histogram}. The left column shows the case with $\epsilon_{\mathrm{pm}}=1$, while the right column corresponds to $\epsilon_{\mathrm{pm}}=2.5$. The first row shows $P(\tau)$ in the presence of static force $F$ only (corresponding to very high flipping rate $\omega$). While for $\epsilon_{\mathrm{pm}}=1$, $P(\tau)$ is nearly Gaussian centered at $\tau_0$, for $\epsilon_{\mathrm{pm}}=2.5$ the distribution is an exponential. This indicates that strong polymer-pore interactions make translocation an activated process, where the chain has to surmount the final free-energy barrier before it can completely translocate to the {\it trans} side (cf. Fig.~\ref{fig:free_energy_schema}). For $\epsilon_{\mathrm{pm}}=1$, as $\omega$ decreases, $P(\tau)$ gradually splits into two Gaussian distributions, centered at $\frac{F}{F+A_d}\tau_{0}$ and $\frac{F}{F-A}\tau_{0}$. For $\epsilon_{\mathrm{pm}}=2.5$, as $\omega$ decreases, $P(\tau)$ is changed in a nontrivial way: at intermediate flipping rate $\omega\approx 1/\tau_{0}$, the tail of $P(\tau)$ is shortened, but for lower $\omega$, $P(\tau)$ develops a long tail. The behavior of $P(\tau)$ at intermediate $\omega$ is closely related to the probability $P_{\tau}(\omega)$ in Fig.~\ref{fig:tau_probability_eps25}. Although $f(0)$ is either positive or negative with similar probability, most of the successful translocations finish with $f(\tau)=+A_{d}$, which results in a shorter translocation time. This is the reason for the small probability of long translocation times. On the other hand, at very low $\omega$, $P(\tau)$ becomes a combination of two exponential distributions, each corresponding to the translocation time with the driving force either $F+A_{d}$ or $F-A_{d}$, which results in sharp increase of $\tau(\omega)$.

\subsubsection{Dependence on the chain length $N$}

\begin{figure}
\includegraphics[bb=0.0cm 0.0cm 8.5cm 6.0cm]{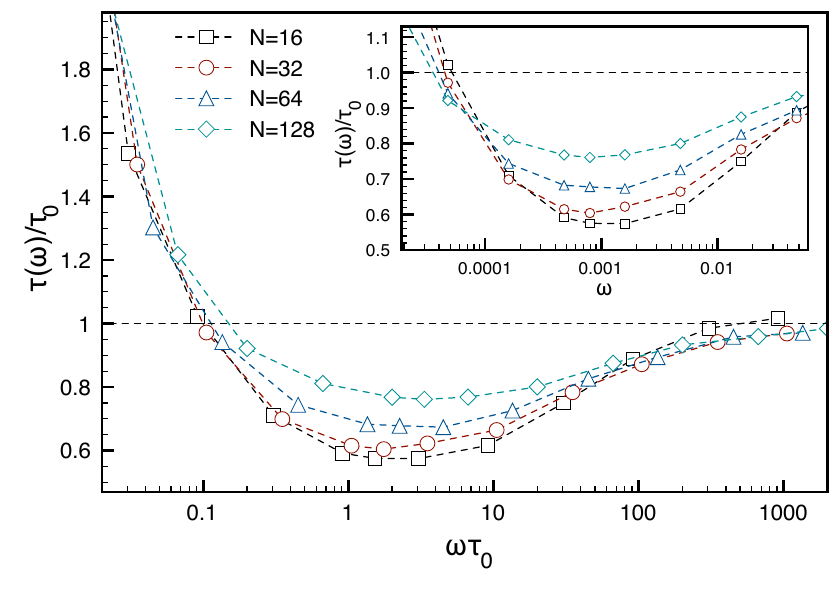}
\caption{Translocation times for chain lengths $16 \leq N \leq 128$ with the dichotomic force and attractive pore. $F=0.5$, $A_{d}=0.2$ and $\epsilon_{\mathrm{pm}}=2.5$. While the optimal rescaled flipping rate ($\omega\tau_0$) shows a slight dependence on $N$ (main figure), the unnormalized flipping rate ($\omega$) is independent of $N$ (inset). The statistical error is smaller than the symbol size.  \label{fig:N_dependence}}
\end{figure}

In the previous section, we found that for large $\epsilon_{\mathrm{pm}}$, the translocation time $\tau$ has a resonant minimum as a function of the flipping rate $\omega$. Here we study how this behavior changes with chain length $N$. Figure~\ref{fig:N_dependence} shows the translocation times $\tau(\omega)$ for $16 \leq N \leq128$, with $\epsilon_{\mathrm{pm}}=2.5$, $F=0.5$, and $A_{d}=0.2$. The optimal flipping rate that yields the minimum of translocation time is roughly independent of $N$ (see the inset of Fig.~\ref{fig:N_dependence}). Since the resonant behavior occurs during the last emptying process, this indicates that $\tau_{3}$ is independent or only weakly depends on the chain length. The free energy barrier of the last emptying process can be approximated as $\Delta F=L(\epsilon_{\mathrm{pm}}-F/2-g(N))$~\cite{muthukumar2003,luo2007a,luo2008b}. Here the first term accounts for the polymer-pore interactions, the second term for the potential energy difference across the membrane due to the driving force, and the last term is due to the entropic free energy~\cite{sung1996,muthukumar1999,muthukumar2003}.  For the pore-emptying process the entropic force $g(N)$ is in the positive direction, slowly increasing with $N$ and eventually saturating for very long chains~\cite{luo2008b}. On the other hand, as shown in Ref.~\cite{luo2008b} for the static driving force and $N\leq 200$, $\tau_{1,2}$ approximately increases as $\tau_{1,2}\sim N^{1.5}$. The normalized translocation time is 
\begin{equation}
\frac{\tau(\omega)}{\tau_{0}}=\frac{\tau_{1,2}(\omega) + \tau_{3}(\omega)}{\tau_{1,2}(0)+ \tau_{3}(0)}=\frac{\tau_{1,2}(\omega)/\tau_{3}(0) + \tau_{3}(\omega)/\tau_{3}(0)}{\tau_{1,2}(0)/\tau_{3}(0) + 1},
\end{equation}
where $\tau_{i}(0)$ is time for the $i$th process in the absence of the time-dependent driving force $f(t)$. In the short chain limit, $\tau_{1,2,}(\omega)\ll\tau_{3}(\omega)$, so that the normalized translocation time is $\frac{\tau(\omega)}{\tau_{0}}\approx \frac{\tau_{3}(\omega)}{\tau_{3}(0)}$.
On the other hand, in the long chain limit, $\tau_{1,2}(\omega)\gg\tau_{3}(\omega)$, giving $\frac{\tau(\omega)}{\tau_{0}}\approx \frac{\tau_{1,2}(\omega)}{\tau_{1,2}(0)}$. These limiting situations predict that for short chains, one should observe a strong minimum in $\tau(\omega)$, whereas for very long chains, the minimum should vanish. This trend can be observed in Fig.~\ref{fig:N_dependence}, where the minimum of translocation time becomes less pronounced as $N$ increases. In addition, the optimal flipping rates are quite independent of $N$. This is in contrast to the results of Ref.~\cite{park1998}, where the authors consider the translocation of a rigid rod in the presence of a reflecting boundary condition at $s=0$ (see Fig.~\ref{fig:free_energy_schema}). In that case, all the segments of the polymer are subject to the external forces, making the translocation time very sensitive to their minute changes and the optimal flipping rate decreases with $N$. However, in the present case the number of segments within the attractive pore remains small throughout the translocation process. Thus, the effect of the external forces becomes small as the chain gets longer.

\subsubsection{Dependence on the driving force $F$}

\begin{figure}
\includegraphics[bb=0.0cm 0.0cm 8.5cm 6.0cm]{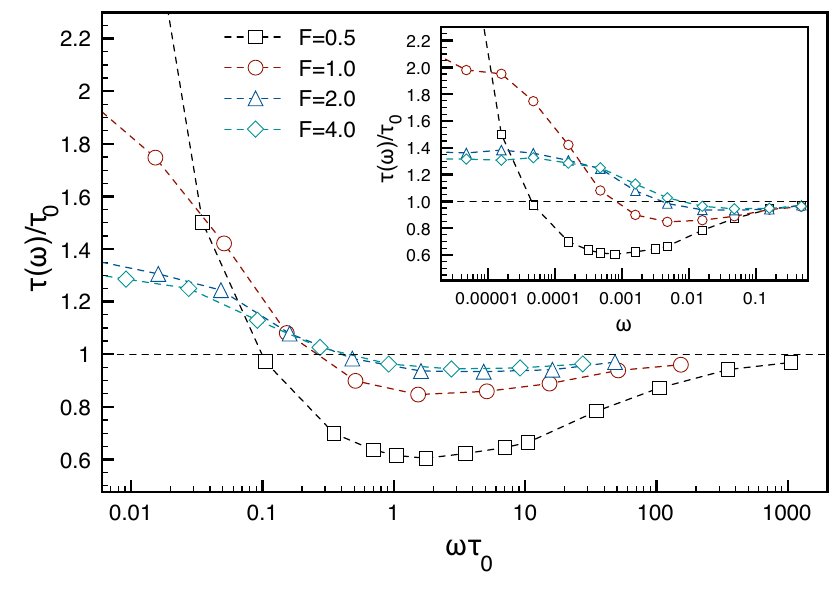}
\caption{Translocation times for driving forces $0.5 \leq F \leq 4$ for the dichotomic force and attractive pore. $A_{d}=0.4F$, $\epsilon_{\mathrm{pm}}=2.5$. In this case, the optimal rescaled flipping rate ($\omega\tau_0$) is roughly independent of $F$ (main figure), while the unnormalized flipping rate ($\omega$) strongly increases with $F$ (inset). The statistical error is smaller than the symbol size. 
\label{fig:force_dependence}}
\end{figure}

Next, we study the effect of changing the driving force magnitude. We consider static driving forces between $0.5 \leq F \leq 4$ with the amplitude $A_d$ fixed as $A_{d}=0.4 F$. The results are shown in Fig.~\ref{fig:force_dependence}. One can see that the optimal flipping rate for the resonant minimum translocation time increases with increasing $F$. In addition, the resonant minimum becomes shallower, finally disappearing for large $F$ and $\tau(\omega)$ becomes a monotonic function of $\omega$. It is because the free energy barrier of the last emptying process vanishes for large $F$. It is of interest to study the critical driving force $F_{c}$, for which $\tau(\omega)$ changes from non-monotonic to monotonic. $F_c$ can be approximated from the condition $\Delta F=0$, so that $F_{c}=2(\epsilon_{\mathrm{pm}}-g(N))$. For $N=32$ and $F=0.5$, $\tau(\omega)$ becomes monotonic for $\epsilon_\mathrm{pm}\lessapprox 1$ (cf. Fig.~\ref{fig:tau_probability_eps1}), giving the estimate $g(N)\approx1$. Therefore, $F_{c}\approx3$ for $\epsilon_{\mathrm{pm}}=2.5$ and $N=32$. This estimate seems to be reasonable as shown in Fig.~\ref{fig:force_dependence}. This result also shows that the condition of the non-monotonic behavior of the translocation time is determined by the competition of  the polymer-pore interaction $\epsilon_{\mathrm{pm}}$ and the driving force $F$.

\subsubsection{Dependence on the driving force amplitude $A_d$}

\begin{figure}
\includegraphics[bb=0.0cm 0.0cm 8.5cm 6.0cm]{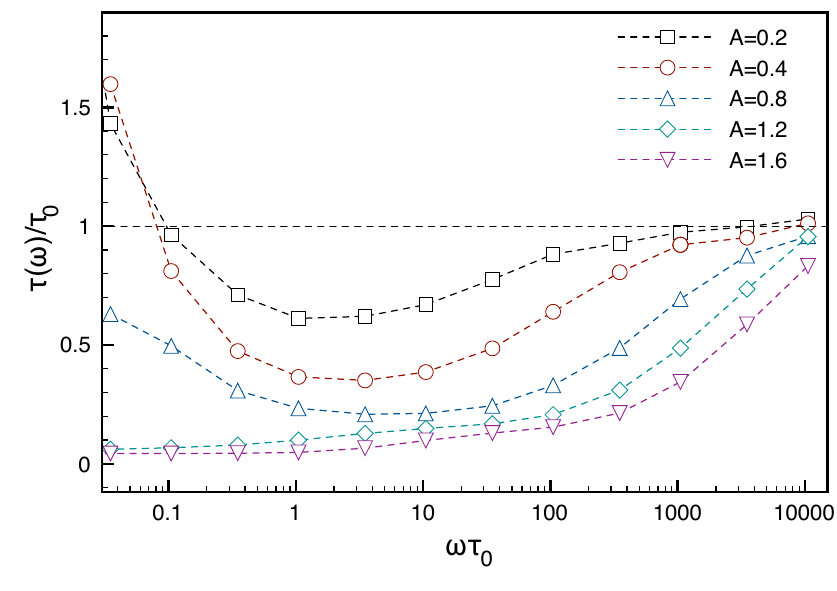}
\caption{Translocation times for the dichotomic force and attractive pore for amplitudes $A_d \in \{ 0.2, 0.4, 0.8, 1.6\}$. $F=0.5$,  $\epsilon_{\mathrm{pm}}=2.5$. The statistical error is smaller than the symbol size.  
\label{fig:tau_A_dichotomic}}
\end{figure}

As the last case of the dichotomic force, we study the effect of changing the dichotomic force amplitude $A_d$ while keeping the static driving force $F$ constant. The results for the translocation time are shown in Fig.~\ref{fig:tau_A_dichotomic}. With increasing $A_d$, the resonant minimum becomes deeper and the resonance flipping rate $\omega$ gradually increases. For very large $A_d$, the resonance disappears and the translocation time becomes a monotonic function of $\omega$, similar to the non-attractive pore case. This shift in behavior is because the selectivity of initial sign of $f(0)$ becomes stronger for larger $A_d$. The transition to this regime happens when the initial barrier (see Fig.~\ref{fig:free_energy_schema}) that prevents the chain escape to the \textit{cis} side becomes comparable to the thermal energy and the escapes become frequent. For the negative dichotomic force, $f(0)=-A_d$, the barrier can be written in the form $\Delta F_\mathrm{cis}=\epsilon_{\mathrm{pm}}+(F-A_d)/2-g(N)$. The pore length $L$ does not enter the relation because in the initial configuration, only the first bead is inside the pore. For $\epsilon_\mathrm{pm}=2.5$ and $k_BT=1.2$, the requirement $\Delta F_\mathrm{cis}\approx k_BT$ gives the estimate $A_d\approx 1$ for the transition from the nonmonotonic $\tau(\omega)$ to the monotonic one. This estimate matches the data in Fig.~\ref{fig:tau_A_dichotomic}.

\subsection{Attractive pore, periodic driving force}

For the attractive pore, the dichotomic and periodic driving forces give qualitatively very similar results. Also for the periodic driving force, $f(t)=A\sin(\omega t + \phi)$, a resonant minimum of the average translocation time appears, in contrast to the non-attractive pore case. In addition, the location (frequency $\omega$) and the depth of the minimum depends on the various parameters in essentially the same way as described in the previous section, which indicates that the origin of the resonance is the same: the time-dependent force being most co-operative to translocation during the pore-emptying time $\tau_3$. However, there are also some obvious differences. For the periodic driving force, oscillatory behavior similar to the one described in Section~\ref{sec:nonattr_periodic} emerges, in addition to the resonant activation. In this Section, we will briefly describe the essential differences between the two driving schemes and examine some of the implications of employing the periodic driving force.

\subsubsection{Dependence on the polymer-pore interaction strength $\epsilon_{\mathrm{pm}}$}

\begin{figure}
\includegraphics[bb=0.0cm 0.0cm 8.5cm 6.0cm]{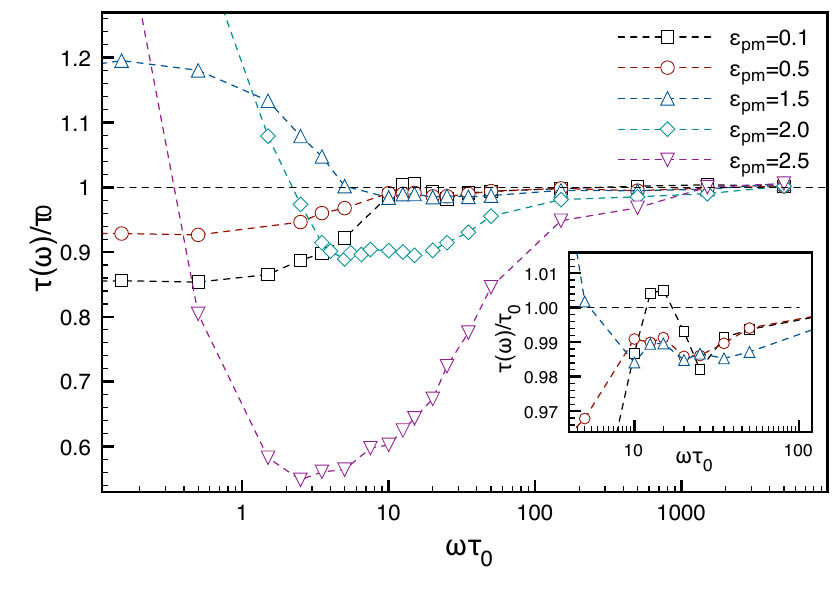}
\caption{Translocation time $\tau$ as a function of frequency $\omega$ for the periodic driving force $f(t)=A\sin(\omega t +\phi)$ for $0.1 \leq \epsilon_\mathrm{pm} \leq 2.5$. $F=0.5$, $A=0.3$ and $N=32$. The inset shows a magnification of the data for $0.1 \leq \epsilon_\mathrm{pm} \leq 1.5$. The statistical error is smaller than the symbol size.  \label{fig:tau_diff_eps}}
\end{figure}

First, it is instructive to consider the dependence of the translocation time $\tau(\omega)$ on the strength of the polymer-pore interaction strength $\epsilon_\mathrm{pm}$. As shown in Fig.~\ref{fig:tau_diff_eps}, for low interaction strengths, one recovers the transition from fast to slow translocation with local minima and maxima in $\tau(\omega)$, characteristic of the non-attractive pore case. For larger $\epsilon_\mathrm{pm}$, a resonant minimum in $\tau(\omega)$ develops, similarly to the dichotomic driving force. However, the global maximum observed for the non-attractive pore persists, although it is reduced to a local maximum located within the resonance minimum. This local maximum arises because of the interplay of the periodic forcing and the non-uniform distribution of the phase $\phi$, as we will discuss below.

\subsubsection{Dependence on the driving force amplitude $A$}

\begin{figure}
\includegraphics[bb=0.0cm 0.0cm 8.5cm 6.0cm]{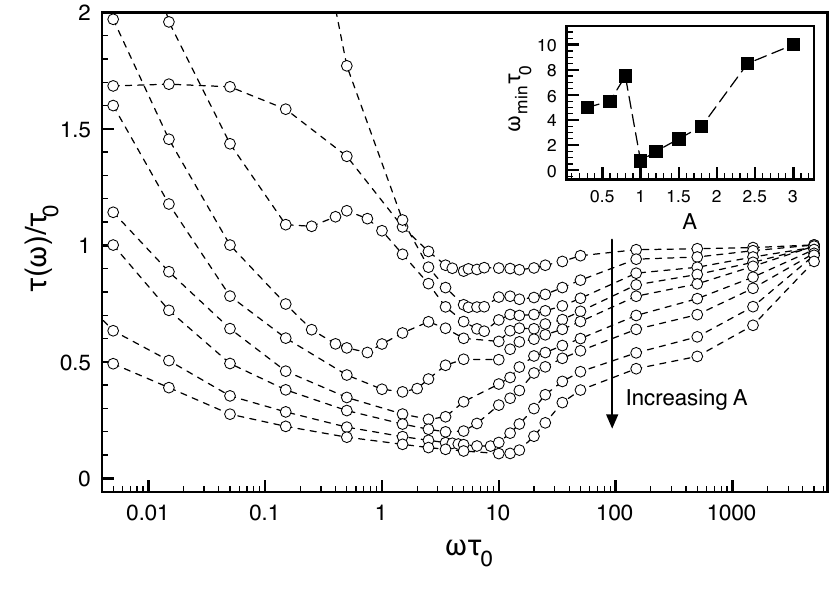}
\caption{Translocation time $\tau$ as a function of frequency $\omega$ for the periodic driving force $f(t)=A\sin(\omega t +\phi)$ for amplitudes $A\in \{ 0.3, 0.6, 0.8, 1.0, 1.2, 1.5, 1.8, 2.4, 3.0 \}$. Other parameters are $F=0.5$, $\epsilon_\mathrm{pm}=2.0$ and $N=32$. The inset shows the dependence of the frequency  $\omega_\mathrm{min}$ of the global minimum translocation time on the amplitude $A$. Here $\tau_0\approx 500 \pm 6$.  The statistical error is smaller than the symbol size. \label{fig:tau_attr_per_diffA}}
\end{figure}

\begin{figure}
\includegraphics[bb=0.0cm 0.0cm 8.5cm 6.0cm]{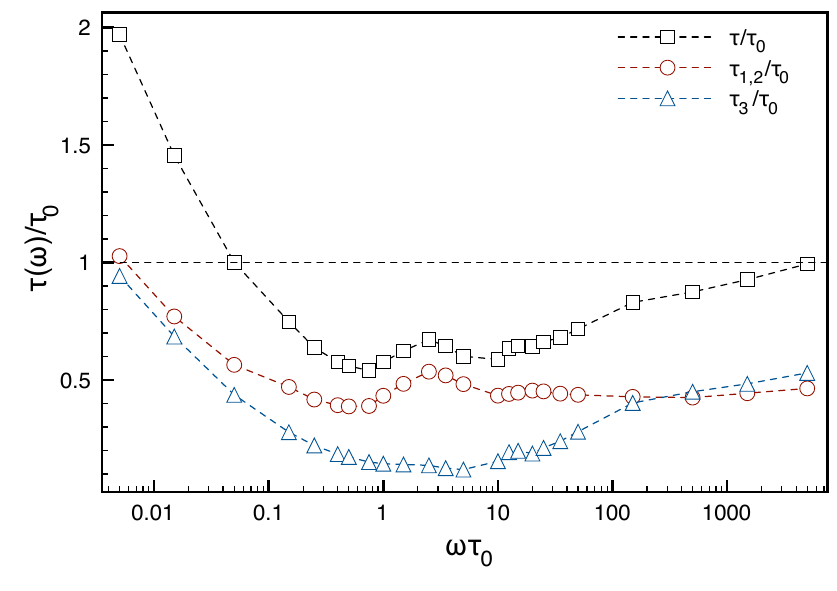}
\caption{The translocation time $\tau$ and its components $\tau_{1,2}$ and $\tau_3$ as a function of $\omega$, showing that the leftmost minimum in $\tau(\omega)$ is associated with $\tau_{1,2}$. Parameter values used are $F=0.5$, $A=1.0$, $\epsilon_\mathrm{pm}=2.0$ and $N=32$. The statistical error is smaller than the symbol size.  \label{fig:tau_attr_per_A10}}
\end{figure}

To highlight the differences between the sinusoidal and dichotomic driving forces, we look at how the translocation time $\tau(\omega)$ changes with the driving force amplitude $A$.  For the dichotomic force, as $A_d$ is increased, one merely crosses from the non-monotonic $\tau(\omega)$ with the resonant minimum to the monotonic $\tau(\omega)$ characteristic of the non-attractive pore case. The sinusoidal driving force, on the other hand, exhibits much richer behavior. As shown in Fig.~\ref{fig:tau_attr_per_diffA}, as the amplitude $A$ is increased, the minimum becomes deeper and slowly moves toward higher frequencies. In addition to the original one, another (local) minimum appears at the low-frequency end of the spectrum and travels down the $\tau(\omega)$ curve as $A$ is increased. Eventually, the new minimum becomes a global one. This produces a sudden transition in the frequency of minimum translocation time, $\omega_\mathrm{min}$, as shown in the inset of Fig.~\ref{fig:tau_attr_per_diffA}. Finally, at sufficiently large $A$, the new minimum merges with the original one. To better understand this complex behavior, we again divide the translocation time $\tau$ to the three components $\tau_1$, $\tau_2$ and $\tau_3$.  Looking at $\tau_{1,2}\equiv\tau_1+\tau_2$ and $\tau_3$ separately reveals that the original global minimum of $\tau(\omega)$ is associated with the pore emptying time $\tau_3$, as shown in Fig.~\ref{fig:tau_attr_per_A10}. The additional minimum, on the other hand, is related to the periodic back-and-forth movement of the chain, which is visible as a nonmonotonic behavior in $\tau_{1,2}(\omega)$. 

\begin{figure}
\includegraphics[bb=0.0cm 0.0cm 8.5cm 6.6cm]{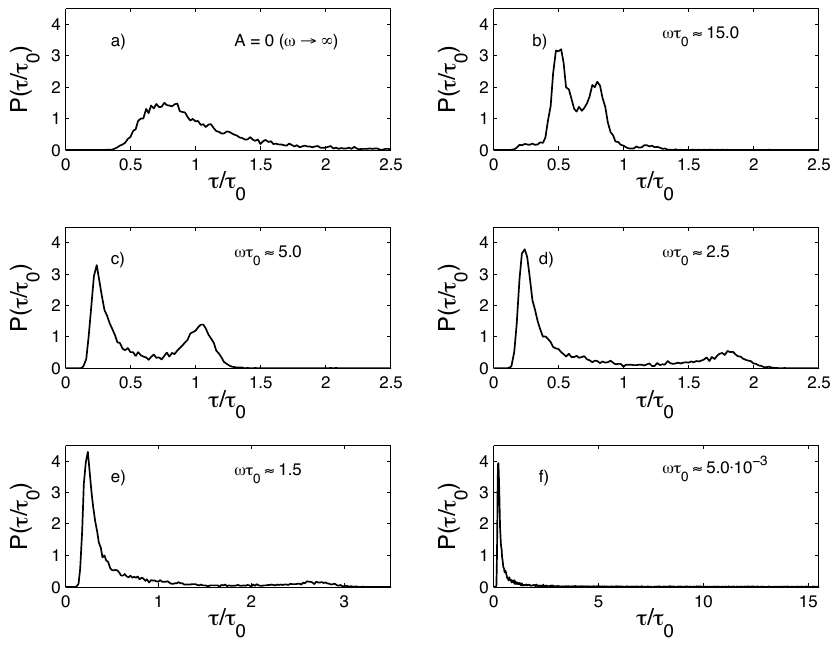}
\caption{Distribution of translocation times for the periodic driving force with $N=32$, $\epsilon_\mathrm{pm}=2.0$, $F=0.5$, $A=0.0$ (panel a), $A=1.0$ (panels b--f). \label{fig:tau_hist_attr_per_A10}}
\end{figure}

Let us first examine the pore emptying time $\tau_3$, since that shares many similarities with the dichotomic force case. In both cases, the minimum of $\tau_3(\omega)$ occurs due to resonant activation. At the corresponding resonant frequency, the probability $P_\tau$ also reaches a maximum (not shown), similarly to the dichotomic force case. As shown in Fig.~\ref{fig:tau_attr_per_A10}, for the sinusoidal force, there is also a small local maximum in $\tau_3(\omega)$ at $\omega\tau_0\approx 15$. Surprisingly, here $P_\tau$ also has a local maximum, which should indicate efficient crossing of the final barrier. However, instead of the expected decrease in $\tau_3$, one sees a slight increase. The reason is that there is a special mismatch between the frequency $\omega$ and the translocation time $\tau_{1,2}$ so that the period $T_\Omega \equiv 2\pi/\omega \approx \tau_{1,2}$. In other words, typically it takes the chain one period of $f(t)$ just to traverse from its initial position to the configuration where it may try to surmount the final free-energy barrier (cf. Fig.~\ref{fig:free_energy_schema}). Thus, the chain essentially misses the first opportunity to cross the final barrier, which slightly increases $\tau_3$. This can be also seen as a suppressed first peak in the translocation time distribution of Fig.~\ref{fig:tau_hist_attr_per_A10}(b).

Finally, let us look at the translocation times $\tau_1$ and $\tau_2$. The combined time $\tau_{1,2}$ shows features similar to the non-attractive pore case, where the periodic driving force produces a series of alternating minima and maxima. In the case of the attractive pore, the selection over the initial phase is weaker because of the free-energy barrier that prevents the escape to the {\it cis} side. Consequently, the qualitative behavior of the $\tau_{1,2}(\omega)$ curve is closer to the model of Eq.~(\ref{eq:toy_model}) with uniformly distributed phase (see Fig.~\ref{fig:nonattr_tau_different_models}). Essentially, the local maximum in $\tau_{1,2}$ is produced by the interplay of $\omega$-dependence of the distribution of $\phi$ and the periodicity of the driving force. Close to the resonant minimum, $\omega\tau_0\approx 5$, the distribution is bimodal, as shown in Fig.~\ref{fig:tau_hist_attr_per_A10}. Similarly to the non-attractive pore case, the first peak corresponds to the events that occur within the first half-period of the force $f(t)$, i.e., $0< \tau < T_\Omega/2$, and whose initial phase is typically $-\pi/4 < \phi < \pi/2$. Therefore, for these trajectories, $f(t)>0$ for most of the process, and translocation occurs faster than average. In contrast, the second peak corresponds to $\pi/2 < \phi < 7\pi/4$ and $T_\Omega/2< \tau< T_\Omega$. As the frequency $\omega$ decreases, the second peak moves further to the right (Fig.~\ref{fig:tau_hist_attr_per_A10}(d),(e)). This increases the average translocation time. At the same time, the area under the peak decreases, because with decreasing $\omega$, the distribution of $\phi$ becomes less uniform, favoring $\phi$ belonging to the first peak. This tends to decrease $\tau$. The combination of these two factors creates the maximum of $\tau_{1,2}$. For larger $A$, the selection over $\phi$ is stronger, so the second factor starts to dominate already at relatively high frequencies. Conversely, for small $A$, the second peak in $P(\tau)$ persists for even very small $\omega$. This explains why the maximum occurs at lower frequencies for small $A$, and moves towards the high-frequency end as $A$ is increased.

\section{Summary}

In this work, we have studied the translocation of polymers under a time-dependent driving force using Langevin dynamics simulations. In particular, we have extracted the dependence of the average translocation time on the flipping rate $\omega$ of the dichotomic driving force and the corresponding dependence on the angular frequency $\omega$ for the sinusoidal driving force. We have also examined the influence of various other physical parameters on the translocation dynamics.

We find that the interactions between the polymer and the pore play a fundamental role in the dynamics of the system. For the non-attractive interactions, the translocation time shows a cross-over to a faster translocation regime at low flipping rates. For the sinusoidal force, in addition to the cross-over, we observe a series of local minima and maxima, produced by the periodicity of the driving signal. However, in this case we do not observe a global minimum of the translocation time for any finite $\omega$. On the other hand, with attractive polymer-pore interactions, which represent naturally occurring biological pores such as the $\alpha$-hemolysin, the situation is very different. In this case, the translocation becomes a thermally activated process due to the attraction between the pore and the polymer. Optimal modulation of time-dependence driving force induces a resonant activation, manifesting as a global minimum in the translocation time at finite $\omega$. We also find that, although the details of this resonance depend on various system parameters, in general the resonance is quite robust and occurs for both the dichotomic and sinusoidal driving force. Typically  the resonant flipping rate (angular frequency) is found in the neighborhood of $\omega \approx 1/\tau_0$, with $\tau_0$ being the translocation time without the time-dependent component of the driving force.  For an experimentally typical translocation time of the order of  100~$\mu s$~\cite{meller2000}, this corresponds to the rate (frequency) in the kilohertz regime.

Theoretically, the occurrence of the resonance relies on the existence of a free energy well, from which the polymer escapes via thermal activation. In practice, to observe the resonant behavior, one should choose the physical parameters so that one has:  1) strong enough polymer-pore interactions, 2) relatively short chain length, 3) small enough static driving force. For example, a poly(dA)$_{100}$ chain driven through an $\alpha$-hemolysin pore should display the resonance for pore voltages of roughly $V\lessapprox 1000$~mV. For $V\approx 200$~mV, we would expect to find the resonance in the neighborhood of $\omega\approx 1 - 10$ kHz at room temperature. Furthermore, in the case of the dichotomic driving force, one should also have a relatively small amplitude of the time-dependent force, whereas for the sinusoidal force, even significantly larger amplitudes can still produce the resonance. In the latter case, a more complicated behavior emerges, as the driving force not only assists translocation during the pore-emptying time $\tau_3$, but also significantly alters to the overall motion of the chain. Our findings suggest that time-dependent driving forces may play a fundamental part in polymer translocation in biological systems, and may also be useful in practical applications such as sorting and sequencing of DNA molecules.

\acknowledgments

This work has been supported in part by the Academy of Finland through its COMP Center of Excellence and Transpoly Consortium grant, and through the General Individual Research Program as well as BK 21 Program administered by the Korea Research Foundation. TI acknowledges the financial support of the Finnish Doctoral Programme in Computational Sciences (FICS) and the Finnish Foundation for Technology Promotion (TES). The authors also wish to thank CSC, the Finnish IT center for science, for allocation of computer resources.


\end{document}